\documentclass{webofc}
\usepackage[varg]{txfonts}   
\def\planck{{\it Planck}} 
\def\chandra{{\it Chandra}}

\begin{document}

\def\planck{{\it Planck}} 
\def\aj{AJ}%
\def\araa{ARA\&A}%
\def\apj{ApJ}%
\def\apjl{ApJ}%
\def\apjs{ApJS}%
\def\aap{Astron. Astrophys.}%
 \def\aapr{A\&A~Rev.}%
\def\aaps{A\&AS}%
\def\mnras{MNRAS}
\def\ssr{SSRv}
\def\nat{Nature}
\def\jcap{JCAP}
\title{Mapping the intracluster medium temperature in the era of NIKA2 and MUSTANG-2}
%
%

\author{\firstname{F.}~\lastname{Ruppin}\inst{\ref{MIT}}\fnsep\thanks{\email{f.ruppin@ip2i.in2p3.fr}}
  \and \firstname{R.}~\lastname{Adam} \inst{\ref{LLR}}
  \and  \firstname{P.}~\lastname{Ade} \inst{\ref{Cardiff}}
  \and  \firstname{H.}~\lastname{Ajeddig} \inst{\ref{CEA}}
  \and  \firstname{P.}~\lastname{Andr\'e} \inst{\ref{CEA}}
  \and \firstname{E.}~\lastname{Artis} \inst{\ref{LPSC}}
  \and  \firstname{H.}~\lastname{Aussel} \inst{\ref{CEA}}
  \and  \firstname{A.}~\lastname{Beelen} \inst{\ref{IAS}}
  \and  \firstname{A.}~\lastname{Beno\^it} \inst{\ref{Neel}}
  \and  \firstname{S.}~\lastname{Berta} \inst{\ref{IRAMF}}
  \and  \firstname{L.}~\lastname{Bing} \inst{\ref{LAM}}
  \and  \firstname{O.}~\lastname{Bourrion} \inst{\ref{LPSC}}
  \and  \firstname{M.} \lastname{Brodwin} \inst{\ref{UMis}}
  \and  \firstname{M.}~\lastname{Calvo} \inst{\ref{Neel}}
  \and  \firstname{A.}~\lastname{Catalano} \inst{\ref{LPSC}}
  \and  \firstname{B.} \lastname{Decker} \inst{\ref{UMis}}
  \and  \firstname{M.}~\lastname{De~Petris} \inst{\ref{Roma}}
  \and  \firstname{F.-X.}~\lastname{D\'esert} \inst{\ref{IPAG}}
  \and  \firstname{S.}~\lastname{Doyle} \inst{\ref{Cardiff}}
  \and  \firstname{E.~F.~C.}~\lastname{Driessen} \inst{\ref{IRAMF}}
  \and  \firstname{P. R. M.} \lastname{Eisenhardt} \inst{\ref{JPL}}
  \and  \firstname{A.}~\lastname{Gomez} \inst{\ref{CAB}}
  \and  \firstname{A. H.} \lastname{Gonzalez} \inst{\ref{UFlo}}
  \and  \firstname{J.}~\lastname{Goupy} \inst{\ref{Neel}}
  \and  \firstname{F.}~\lastname{K\'eruzor\'e} \inst{\ref{LPSC}}
  \and  \firstname{C.}~\lastname{Kramer} \inst{\ref{IRAME}}
  \and  \firstname{B.}~\lastname{Ladjelate} \inst{\ref{IRAME}}
  \and  \firstname{G.}~\lastname{Lagache} \inst{\ref{LAM}}
  \and  \firstname{S.}~\lastname{Leclercq} \inst{\ref{IRAMF}}
  \and  \firstname{J.-F.}~\lastname{Lestrade} \inst{\ref{LERMA}}
  \and  \firstname{J.-F.}~\lastname{Mac\'ias-P\'erez} \inst{\ref{LPSC}}
  \and  \firstname{A.}~\lastname{Maury} \inst{\ref{CEA}}
  \and  \firstname{P.}~\lastname{Mauskopf} \inst{\ref{Cardiff},\ref{Arizona}}
  \and \firstname{F.}~\lastname{Mayet} \inst{\ref{LPSC}}
  \and  \firstname{M.} \lastname{McDonald} \inst{\ref{MIT}}
  \and  \firstname{A.}~\lastname{Monfardini} \inst{\ref{Neel}}
  \and  \firstname{E.} \lastname{Moravec} \inst{\ref{UFlo}}
  \and  \firstname{M.}~\lastname{Mu\~noz-Echeverr\'ia} \inst{\ref{LPSC}}
  \and  \firstname{L.}~\lastname{Perotto} \inst{\ref{LPSC}}
  \and  \firstname{G.}~\lastname{Pisano} \inst{\ref{Cardiff}}
  \and  \firstname{N.}~\lastname{Ponthieu} \inst{\ref{IPAG}}
  \and  \firstname{V.}~\lastname{Rev\'eret} \inst{\ref{CEA}}
  \and  \firstname{A.~J.}~\lastname{Rigby} \inst{\ref{Cardiff}}
  \and  \firstname{A.}~\lastname{Ritacco} \inst{\ref{IAS}, \ref{ENS}}
  \and  \firstname{C.}~\lastname{Romero} \inst{\ref{Pennsylvanie}}
  \and  \firstname{H.}~\lastname{Roussel} \inst{\ref{IAP}}
  \and  \firstname{K.}~\lastname{Schuster} \inst{\ref{IRAMF}}
  \and  \firstname{S.}~\lastname{Shu} \inst{\ref{Caltech}}
  \and  \firstname{A.}~\lastname{Sievers} \inst{\ref{IRAME}}
  \and  \firstname{S. A.} \lastname{Stanford} \inst{\ref{UCal}}
\and  \firstname{D.} \lastname{Stern} \inst{\ref{JPL}}
  \and  \firstname{C.}~\lastname{Tucker} \inst{\ref{Cardiff}}
  \and  \firstname{R.}~\lastname{Zylka} \inst{\ref{IRAMF}}
}

\institute{
   Kavli Institute for Astrophysics and Space Research, Massachusetts Institute of Technology, Cambridge, MA 02139, USA
  \label{MIT}
  \and
  LLR (Laboratoire Leprince-Ringuet), CNRS, École Polytechnique, Institut Polytechnique de Paris, Palaiseau, France
  \label{LLR}
  \and
  School of Physics and Astronomy, Cardiff University, Queen’s Buildings, The Parade, Cardiff, CF24 3AA, UK 
  \label{Cardiff}
  \and
  AIM, CEA, CNRS, Universit\'e Paris-Saclay, Universit\'e Paris Diderot, Sorbonne Paris Cit\'e, 91191 Gif-sur-Yvette, France
  \label{CEA}
  \and
  Univ. Grenoble Alpes, CNRS, Grenoble INP, LPSC-IN2P3, 53, avenue des Martyrs, 38000 Grenoble, France
  \label{LPSC}
    \and
  Institut d'Astrophysique Spatiale (IAS), CNRS, Universit\'e Paris Sud, Orsay, France
  \label{IAS}
  \and
  Institut N\'eel, CNRS, Universit\'e Grenoble Alpes, France
  \label{Neel}
  \and
  Institut de RadioAstronomie Millim\'etrique (IRAM), Grenoble, France
  \label{IRAMF}
  \and
  Aix Marseille Univ, CNRS, CNES, LAM (Laboratoire d'Astrophysique de Marseille), Marseille, France
  \label{LAM}
  \and 
  Department of Physics and Astronomy, University of Missouri, 5110 Rockhill Road, Kansas City, MO 64110, USA
  \label{UMis}
  \and
  Dipartimento di Fisica, Sapienza Universit\`a di Roma, Piazzale Aldo Moro 5, I-00185 Roma, Italy
  \label{Roma}
  \and
  Univ. Grenoble Alpes, CNRS, IPAG, 38000 Grenoble, France 
  \label{IPAG}
  \and
  Jet Propulsion Laboratory, California Institute of Technology, Pasadena, CA 91109, USA
\label{JPL}
\and
  Centro de Astrobiolog\'ia (CSIC-INTA), Torrej\'on de Ardoz, 28850 Madrid, Spain
  \label{CAB}
  \and  
  Department of Astronomy, University of Florida, 211 Bryant Space Center, Gainesville, FL 32611, USA
\label{UFlo}
\and
  Instituto de Radioastronom\'ia Milim\'etrica (IRAM), Granada, Spain
  \label{IRAME}
  \and 
  LERMA, Observatoire de Paris, PSL Research University, CNRS, Sorbonne Universit\'e, UPMC, 75014 Paris, France  
  \label{LERMA}
  \and
  School of Earth and Space Exploration and Department of Physics, Arizona State University, Tempe, AZ 85287, USA
  \label{Arizona}
  \and 
  Laboratoire de Physique de l’\'Ecole Normale Sup\'erieure, ENS, PSL Research University, CNRS, Sorbonne Universit\'e, Universit\'e de Paris, 75005 Paris, France 
  \label{ENS}
  \and
  Department of Physics and Astronomy, University of Pennsylvania, 209 South 33rd Street, Philadelphia, PA, 19104, USA
  \label{Pennsylvanie}
  \and 
  Institut d'Astrophysique de Paris, Sorbonne Universit\'e, CNRS (UMR 7095), 75014 Paris, France
  \label{IAP}
  \and
  Caltech, Pasadena, CA 91125, USA
  \label{Caltech}
  \and Department of Physics, University of California, One Shields Avenue, Davis, CA 95616, USA
  \label{UCal}
}

\abstract{%
We present preliminary results from an on-going program that aims at mapping the intracluster medium (ICM) temperature of high redshift galaxy clusters from the MaDCoWS sample using a joint analysis of shallow X-ray data obtained by \chandra\ and high angular resolution Sunyaev-Zel'dovich (SZ) observations realized with the NIKA2 and MUSTANG-2 cameras. We also present preliminary results from an on-going Open Time program within the NIKA2 collaboration that aims at mapping the ICM temperature of a galaxy cluster at $z=0.45$ from the resolved detection of the relativistic corrections to the SZ spectrum. These studies demonstrate how high angular resolution SZ observations will play a major role in the coming decade to push the investigation of ICM dynamics and non-gravitational processes to high redshift before the next generation X-ray observatories come into play.
}
\maketitle
\section{Introduction}\label{sec:intro}

Mapping the temperature of the hot plasma within galaxy clusters from the spectroscopic information embedded in X-ray data has enabled tremendous advances in the understanding of merger dynamics,  sloshing,  and AGN feedback at low redshift \citep[\emph{e.g.}][]{rus12,ros13,raf12}.  However,  producing such temperature maps is currently unfeasible for most clusters at $z > 0.5$ because of the limited effective area of current X-ray observatories.  Unfortunately,  the next generation of X-ray observatories such as \emph{Athena} and \emph{Lynx} will not be operational until the 2030s.  In the meantime it is essential to push the investigation of ICM dynamics and thermodynamic properties to higher redshift in order to understand how galaxy clusters formed and evolved. Fortunately,  during the last decade,  we have witnessed the arrival of instruments capable of mapping the SZ effect at high angular resolution on scales of few arcmin \citep[\emph{e.g.}][]{mon10,ada18,dic14}. The amplitude of the SZ signal being redshift independent and proportional to the ICM pressure content, the joint analysis of SZ and X-ray data offers a powerful way of mapping the ICM temperature distribution at high redshift \citep[\emph{e.g.}][]{ada17,rup20}.  Another challenging method that enables mapping the ICM temperature at high redshift consist in detecting the relativistic corrections to the SZ effect (rSZ).  This method has the advantage of being independent of X-ray observations.  It could therefore allow us to unveil potential calibration biases affecting the temperature measurements realized with current X-ray observatories.  In this paper,  we present preliminary results from two on-going programs using respectively joint analyses of X-ray and SZ data and the detection of the rSZ effect in order to map the ICM temperature of high redshift clusters.

\section{Mapping the ICM temperature with joint X-ray / SZ analyses}\label{sec:XSZ}

\begin{figure*}
\centering
\includegraphics[height=4.1cm]{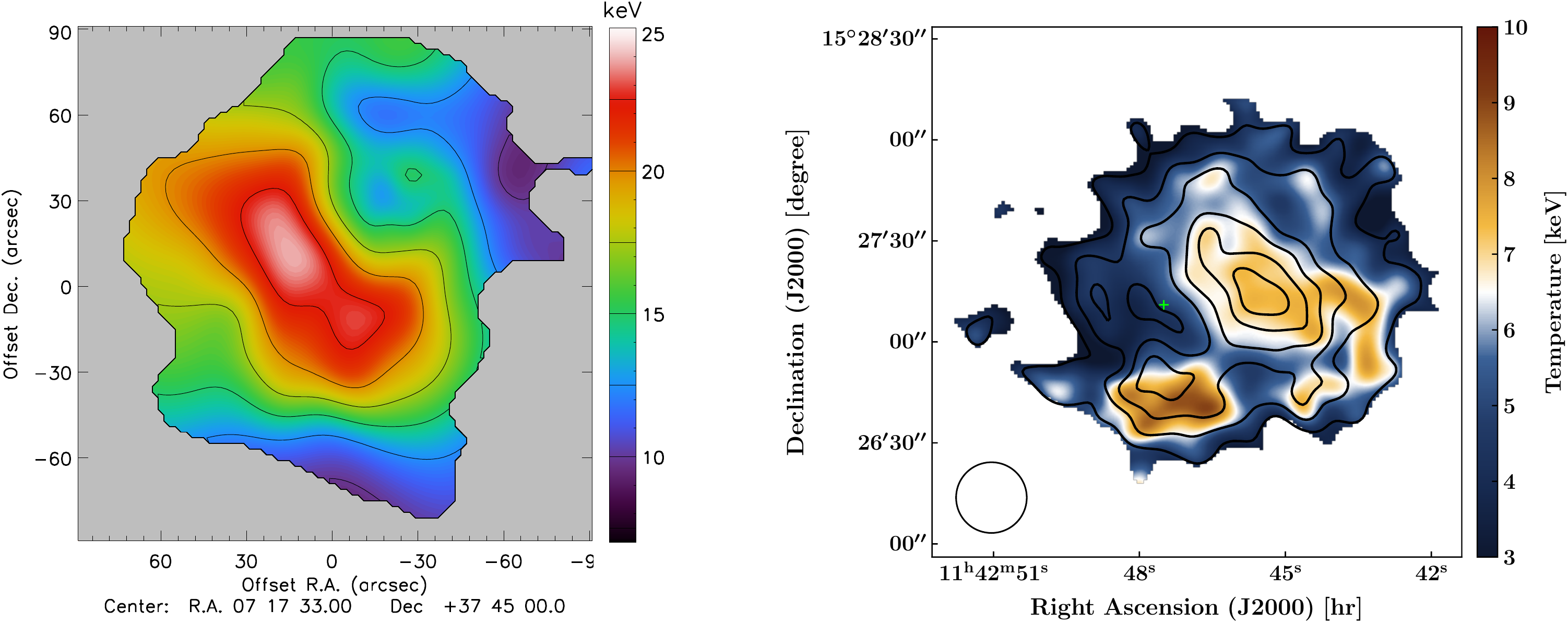}
\caption{{\footnotesize \textbf{Left:}  Spectroscopic temperature map derived from \chandra\ for a cluster at $z = 0.55$. \textbf{Right:} Temperature map obtained from the joint analysis of \chandra\ X-ray and NIKA2 SZ observations for a cluster at $z=1.2$. The relative uncertainties on temperature are similar in both panels. Figures extracted from \citep{ada17,rup20}.}}
\label{fig:T_maps}
\end{figure*}

\begin{figure*}
\centering
\includegraphics[height=4.9cm]{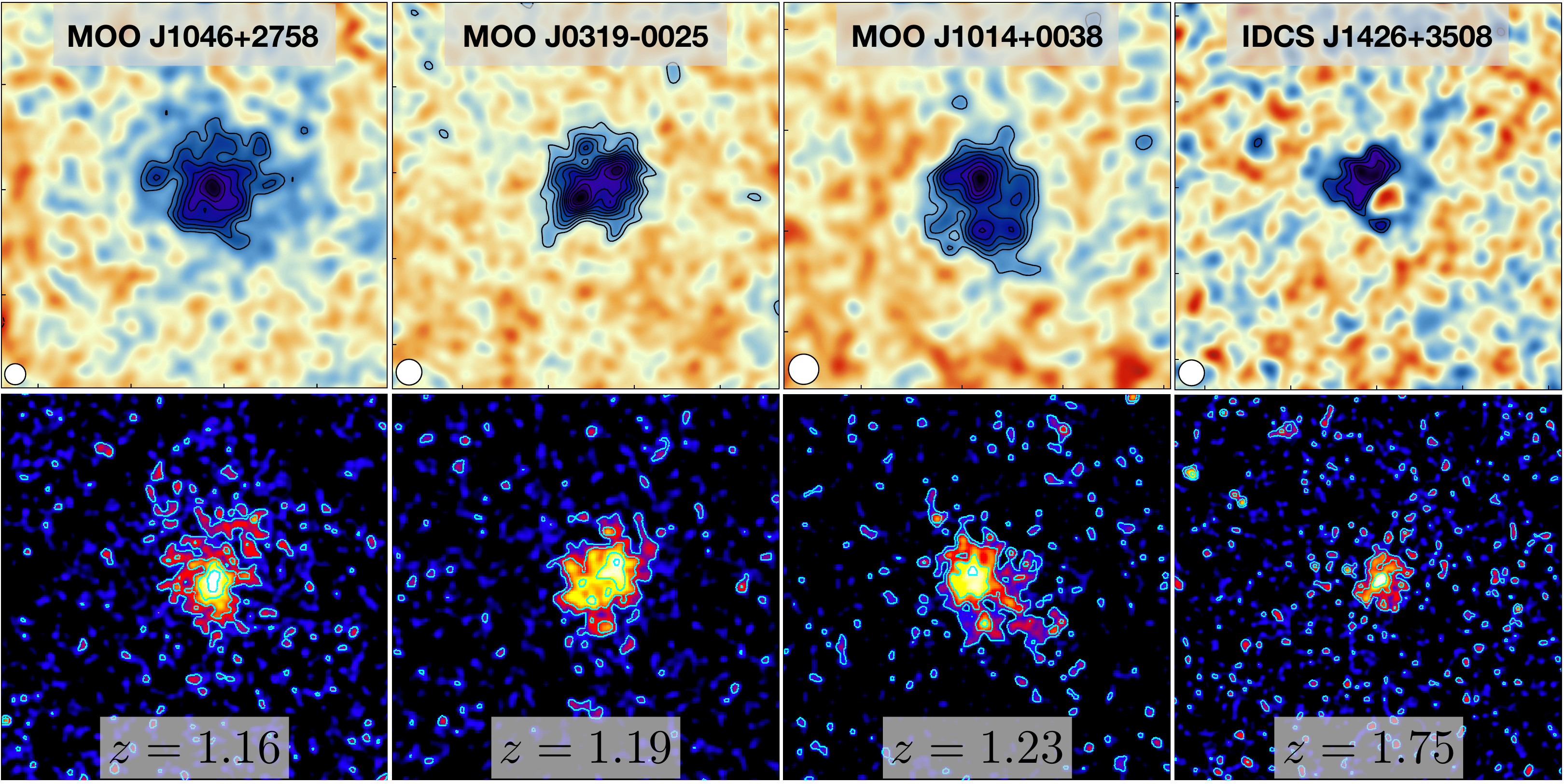}
\caption{{\footnotesize Preliminary SZ (top) and X-ray (bottom) maps obtained with the MUSTANG-2 (far left) and NIKA2 camera at 90~GHz and 150~GHz and the ACIS-I instrument on-board \chandra\ for four clusters in the MOO-X/SZ sample.  Contours in the MUSTANG-2 and NIKA2 maps start at $3\sigma$ with $1\sigma$ steps.  Photon count iso-contours are shown in the X-ray maps for visual purposes. We indicate the name and redshift of each cluster in the upper and lower panels,  respectively. All maps have a side length of about ${\sim}4$~arcmin.}}
\label{fig:mooxsz}
\vspace{-0.6cm}
\end{figure*}

Simulations have shown that the most active period in cluster formation history is expected at redshifts $1 < z < 2$ \citep[\emph{e.g.}][]{poo07}.  Studying merger dynamics and non-gravitational processes such as AGN feedback through temperature mapping thus requires high angular resolution X-ray and SZ observations at $z > 1$. \\
The \emph{Massive and Distant Clusters of WISE Survey} (MaDCoWS) is an Infra-Red (IR) survey that has enabled detecting hundreds of massive clusters at $z > 1$ \citep{gon19} observable from the northern hemisphere,  where the NIKA2 and MUSTANG-2 SZ cameras are operational.  We have selected 10 clusters from the MaDCoWS catalog in order to realize high angular resolution follow-up observations in X-ray and SZ with \chandra, NIKA2,  and MUSTANG-2. This sample,  dubbed MOO-X/SZ,  covers a redshift range from $z=0.9$ to $z=1.75$. The joint analysis of the SZ and X-ray data obtained for each MOO-X/SZ cluster will allow us to investigate ICM physics with a level of details and precision that is not achievable with current X-ray observatories on their own at such high redshifts. \\
We have already completed the analysis of the most massive cluster in this sample, \emph{i.e.} MOO\,J1142$+$1527 at $z=1.2$ \citep{rup20}.  As shown in Fig.~\ref{fig:T_maps},  the quality of the temperature map of this high-$z$ cluster that we realized from the joint analysis of \chandra\ and NIKA2 data with a combined exposure of only 23.2 hours (right) is similar to the one achieved from the analysis of the X-ray spectroscopic data obtained with a 42.5 hour \chandra\ exposure of the $z=0.55$ cluster MACS\,J0717.5$+$3745 (left) \citep{ada17}.  Obtaining a similar map using only X-ray information from \chandra\ at such high redshift would have required an exposure at least 30 times greater.  This highlights the strong complementarity between X-ray and high angular resolution SZ observations to go beyond the characterization of individual systems at $z > 1$ and start investigating the ICM properties in samples of high redshift clusters. \\
Building upon the success of this first study,  a NIKA2 Open Time (OT) program aiming at mapping the SZ signal of 5 MOO-X/SZ clusters already observed by \chandra\ as well as a Large Program (LP) dedicated to the \chandra\ observation of 4 MOO-X/SZ clusters already observed by MUSTANG-2 have been accepted. The NIKA2 OT observations were completed in March 2021 and the \chandra\ LP observations are still on-going.  In Fig.~\ref{fig:mooxsz},  we present preliminary maps of 4 MOO-X/SZ clusters obtained in SZ (top) and X-ray (bottom). The first SZ map on the left has been obtained with MUSTANG-2 at 90~GHz and the three other maps present NIKA2 data at 150~GHz \citep{per20}. The SZ signal is detected at $>7\sigma$ in all maps.  We observe similar cluster morphologies in SZ and X-ray and we identify ICM disturbances in each cluster.  For example,  MOO\,J0319$-$0025 at $z=1.19$ displays a clear bi-modal morphology while MOO\,J1014$+$0038 at $z=1.23$ presents ICM substructures in the south-west region. These preliminary results tend to confirm our expectation that ICM disturbances due to merging events are more frequently observed at $z > 1$ than in low redshift cluster samples such as REXCESS \citep{boh07} or CCCP \citep{hoe12}.  Based on the tools that we developed for the analysis of MOO\,J1142$+$1527, we are currently analyzing jointly our X-ray and SZ data sets in order to provide a detailed characterization of the ICM properties in a sample of $z>0.9$ clusters for the first time. In particular,  we are producing ICM temperature and entropy maps that we cross correlate with the galaxy distribution of each cluster in order to understand their merger dynamics.  We also aim at exploring the connection between the ICM thermodynamic properties of all MOO-X/SZ clusters and the potential radio emission of the brightest cluster galaxy in order to better understand AGN feedback at high redshift.  The analysis of the spectroscopic information embedded within our \chandra\ data using prior information on the ICM temperature obtained from the joint X-ray / SZ analysis will allow us to constrain the metallicity content of these cluster in order to understand metal enrichment at high redshift. Finally,  the hydrostatic masses obtained from the estimated density and pressure profiles of these clusters will be compared to the richness-derived masses obtained with \emph{Spitzer} in order to unveil potential systematic effects on the mass-richness scaling relation derived at high redshift. The full characterization of the MOO-X/SZ sample will thus provide valuable information that will pave the way to an in-depth understanding of cluster formation at $z > 1$ and to precise cluster cosmological constraints from the forthcoming optical/IR surveys such as \emph{Euclid} \citep{euc19} and \emph{Rubin}-LSST \citep{wu21}. 

\section{Mapping the ICM temperature from the detection of the rSZ effect}\label{sec:rSZ}

Assuming that the contributions to the SZ effect from sub-dominant electron populations such as ultra-relativistic electrons accelerated in AGN jets or electrons arising from the annihilation of dark matter are negligible,  the total relative variation of the CMB specific intensity at frequency $\nu$ induced from the SZ effect is given by:
\begin{equation}
\frac{\Delta I_{\nu}}{I_0} = y_{tSZ}\,f(\nu,T_e) + y_{kSZ}\,g(\nu,T_e,v_z),
\end{equation}
where the amplitudes of the thermal and kinetic SZ signals are given by:
\begin{equation}
y_{tSZ} = \frac{k_BT_e}{m_ec^2}\tau~~~\mathrm{and}~~~y_{kSZ} = -\frac{v_z}{c}\tau~~~\mathrm{with}~~~\tau = \sigma_T\int n_e\,dl.
\end{equation}
The SZ signal thus depends on three quantities: the density of ICM electrons $n_e$ along the line of sight $l$, the ICM temperature $T_e$,  and the bulk velocity of the ICM electrons along the line of sight $v_z$. The relativistic corrections inducing variations of the shape of the tSZ and kSZ spectra $f(\nu,T_e)$ and $g(\nu,T_e,v_z)$ are often referred to as the rSZ effect and depend directly on the value of the ICM temperature.  Sampling the total SZ spectrum with at least three independent data points on a large enough frequency range thus enables estimating the ICM temperature independently from any X-ray observation. The detection of the rSZ effect has already been made using millimeter spectroscopy with Z-Spec \citep{zem12} and by stacking analyses \citep[\emph{e.g.}][]{erl18}. These studies enable measuring the mean ICM temperature of galaxy clusters but they were unable to provide resolved detection of the rSZ effect which is a prerequisite to the production of ICM temperature maps. \\
\begin{figure*}
\centering
\includegraphics[height=6.3cm]{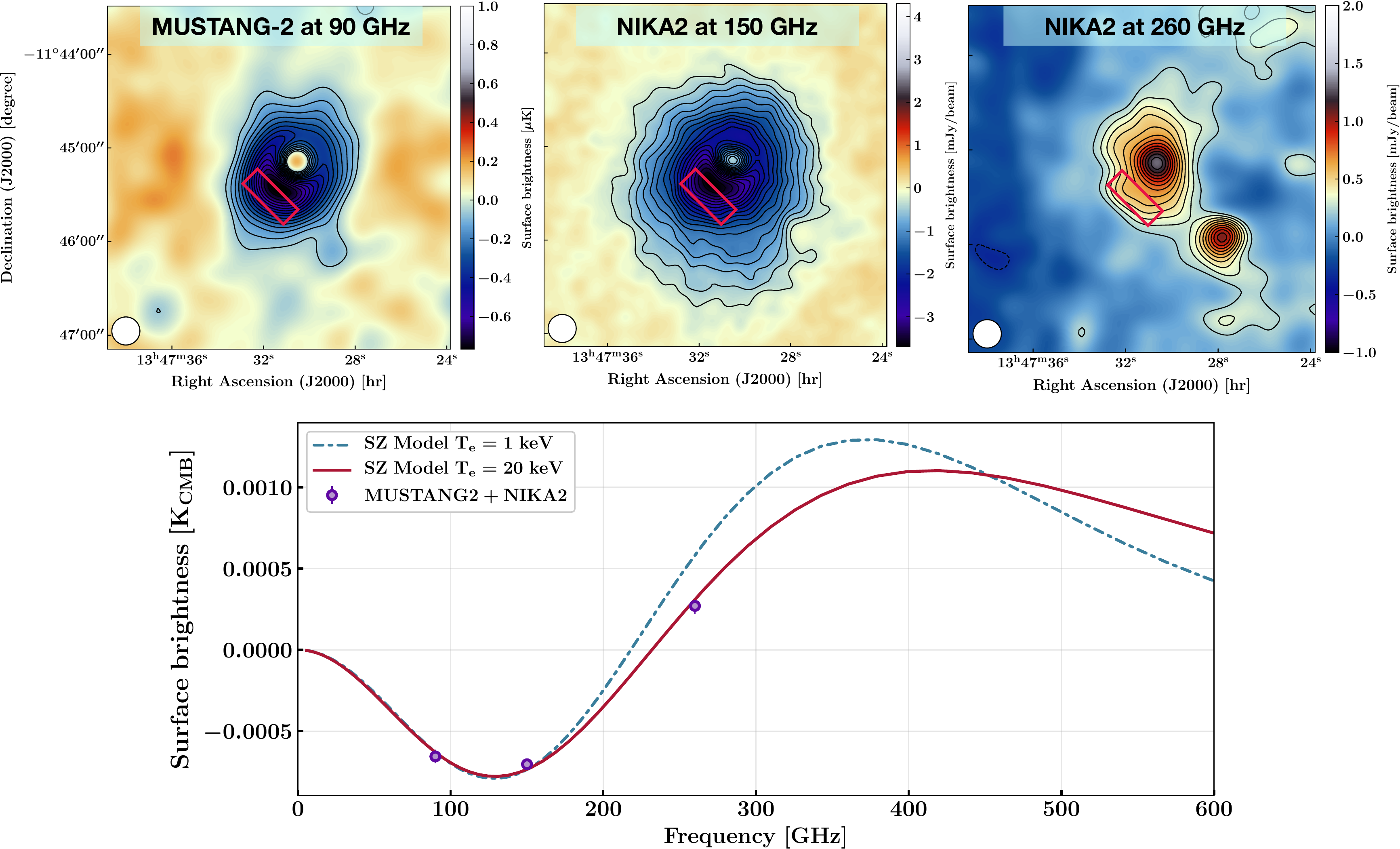}
\caption{{\footnotesize \textbf{Top:} Preliminary SZ maps obtained with MUSTANG-2 and NIKA2 at 90, 150, and 260~GHz,  respectively.  Signal-to-noise contours start at $3\sigma$ with $1\sigma$ increments.  We highlight the region considered for the SZ surface brightness measurements shown in the bottom panel with red rectangles.  \textbf{Bottom:} Preliminary measurements of the SZ surface brightness in the shock region of the cluster at the three frequencies available with NIKA2 and MUSTANG-2 (purple points) along with the expected tSZ spectra for ICM temperatures of 1 and 20~keV.}}
\label{fig:rSZ}
\vspace{-0.6cm}
\end{figure*}To this end, we proposed to use both MUSTANG-2 and NIKA2 in order to map the SZ signal of the most SZ luminous cluster in the sky,  RX\,J1347$-$1145 at $z=0.45$,  at an angular resolution $<18$~arcsec and on scales of few arcmin at 90, 150, and 260~GHz. The MUSTANG-2 observations were realized at the Green Bank Telescope during the instrument commissioning phase for 2~hours. The NIKA2 observations have been conducted in the context of an Open Time program for a total of 24 hours. The top panels of Fig.~\ref{fig:rSZ} show the MUSTANG-2 map (left) obtained with the MIDAS software \citep{rom20} and the NIKA2 maps (middle and right) realized from a preliminary analysis based on the procedure described in \citep{rup20}.  All three maps have been smoothed at an effective resolution of 20~arcsec. The extended SZ signal is detected at $>3\sigma$ on a sky area of about $2\times 1.5~\mathrm{arcmin}^2$. We detect a strong positive signal in the cluster core at 260~GHz due to a radio AGN. This signal compensates the negative SZ signal at 90~GHz and 150~GHz. We also detect a sub-millimeter galaxy to the south-west of the cluster, which also alters its morphology at 150~GHz. As a preliminary test, we measured the average SZ surface brightness in a region (red rectangle in Fig.~\ref{fig:rSZ}) where a shock has been detected in previous analyses \citep{ale02} and far enough from the detected point sources to minimize their contamination.  The measurements are shown as purple points in the lower panel of Fig.~\ref{fig:rSZ}. We use the SZpack software \citep{chl12} in order to compute the expected SZ spectrum for ICM temperatures of 1~keV (dashed blue line) and 20~keV (red line),  assuming a null bulk velocity along the line of sight\footnote{Note that having three SZ constraints will also enable fitting the kinetic SZ component in the total SZ signal.}.  While changing the value of the Compton parameter $y_{tSZ}$ allows us to fit the data at 90~GHz and 150~GHz in both cases,  we are able to exclude an ICM temperature of 1~keV at more than $5\sigma$ in the shock region based on the measurement realized at 260~GHz.  A careful characterization of the point source, cosmic infrared background, and kSZ contamination using \emph{Herschel} and spectroscopic data will allow us to produce the first ICM temperature map of a galaxy cluster based on the resolved detection of the rSZ effect.

\section{Conclusions}\label{sec:conclu}

We have completed the observations of two independent SZ programs aiming at mapping the ICM temperature of high redshift galaxy clusters. The joint analysis of our SZ data and \chandra\ observations will allow us to characterize the ICM properties of a sample of MaDCoWS clusters at $z > 1$ with a level of details that cannot be achieved with current X-ray observatories on their own.  Furthermore,  the resolved detection of the rSZ effect with MUSTANG-2 and NIKA2 will enable providing the first ICM temperature map of a merging cluster without considering any X-ray information.

\section*{Acknowledgements}
\small{We would like to thank the IRAM staff for their support during the campaigns. The NIKA2 dilution cryostat has been designed and built at the Institut N\'eel. In particular, we acknowledge the crucial contribution of the Cryogenics Group, and in particular Gregory Garde, Henri Rodenas, Jean Paul Leggeri, Philippe Camus. This work has been partially funded by the Foundation Nanoscience Grenoble and the LabEx FOCUS ANR-11-LABX-0013. This work is supported by the French National Research Agency under the contracts "MKIDS", "NIKA" and ANR-15-CE31-0017 and in the framework of the "Investissements d’avenir” program (ANR-15-IDEX-02). This work has benefited from the support of the European Research Council Advanced Grant ORISTARS under the European Union's Seventh Framework Programme (Grant Agreement no. 291294). F.R. acknowledges financial supports provided by NASA through SAO Award Number SV2-82023 issued by the Chandra X-Ray Observatory Center, which is operated by the Smithsonian Astrophysical Observatory for and on behalf of NASA under contract NAS8-03060.}

\end{document}